\documentstyle[preprint,aps]{revtex}
\input{psfig}
\begin{document}
\preprint{SBRHI-99-2}


\title{Rescattering of Vector Meson Daughters in High 
	Energy Heavy Ion Collisions}
\author{Stephen C. Johnson\footnote{present address: Lawrence
	Livermore National Laboratory L-305, Livermore, CA 94550},
	Barbara V. Jacak, and Axel Drees}
\address{The University at Stony Brook, Stony Brook, NY  11794-3800}

\maketitle

\begin{abstract}

We consider the role of hadronic rescattering of daughter kaons on the
observed mass spectra from $\phi$ meson decays in ultra-relativistic
heavy ion collisions.  A hadronic cascade code (RQMD v2.4)
shows that $\sim$26\% of all $\phi$'s decaying to
$K^+K^-$ in central Pb+Pb collisions at SPS energies ($E_{beam} = 158
GeV/A$) have a rescattered or absorbed daughter.  This significantly
affects the
reconstructed invariant mass of the pair and shifts
$\phi$ mesons out of the mass peak.  Kaon rescattering depletes the
low velocity region, hardening and broadening the
observed phi $m_t$ and rapidity distributions
respectively, relative to the dilepton channel.  This effect produces an
apparent change in the experimentally determined branching ratio
not necessarily related to chiral symmetry restoration.  Comparisons
to recent experimental measures at CERN energies reveal a possible
mechanism to account for the shape of the observed spectra, though not
their absolute relative magnitude.

\end{abstract}


\subsection{Introduction}

Chiral symmetry, the symmetry which acts separately on left- and right-handed
fields in QCD, is spontaneously broken in nature, resulting in 
associated
Goldstone bosons in the pion field as well as the observed hadron mass
spectra.  
However, it is expected that the spontaneously broken part of chiral 
symmetry may be restored at high temperatures and densities
\cite{koch}.  Such conditions might be reached
in the interior of neutron stars or
during a relativistic heavy ion collision.
In fact, a primary goal of high energy nuclear 
collisions is the creation of such matter and the study of its
properties.

One expected signal of the restoration of chiral
symmetry is a change of the vector meson properties \cite{koch}.
In the nuclear system, the self energy of hadrons is changed by the medium
they inhabit.  A change in
the effective mass of daughter
particles can change the effective lifetime, and consequently the observed
width, of the parent particle that decays in medium.  Such modifications
could be accompanied by a
measureable change in the branching ratio 
of mesons decaying in the medium.  The light vector
mesons ($\rho$, $\omega$, $\phi$)
offer an especially promising channel for study of chiral symmetry
effects due to their multi-channel decays and lifetimes comparable to
the space-time extent of the system produced in heavy ion collisions.
The $\phi$ is of particular interest as the sum
of the daughter mass of its di-kaon decay channel is very close to the mass 
of the
$\phi$.  As a result, even a small change in daughter or parent masses
could measureably alter the decay channel \cite{shuryak}.

Study of vector mesons through their leptonic decay
channels, either $e^+e^-$ or $\mu^+\mu^-$ should provide the cleanest
signal of the changing masses as leptons interact with the nuclear medium
predominantly electromagnetically.  In contrast, decays to hadrons are
affected by strong final state interactions.  If a $\phi$ decays in the 
center of the reaction volume in a heavy ion collision, the lifetime
of the kaon travelling a
distance $d$ through a length $L$ of hadronic matter is
$exp(-Ld\sigma)$.  Approximating the heavy ion collision as a pion bath
with density $d = .5$/fm$^3$ and cross section $\sigma = 10-100$mb,
the 1/e
pathlength of a kaons would be between .2fm and 2fm, much smaller than
the size of the
collision region.  However, while the hadronic daughters interact in the 
medium, leptons should escape unscathed.

The rescattering of hadronic daughters
in the nuclear medium could mimic or obscure effects of chiral symmetry
restoration by causing the
reconstructed invariant mass to fall outside the vector meson peak, 
effectively decreasing the measured yields.  While studies of the
viability of the di-kaon channel of the $\phi$ for studying chiral
symmetry restoration have been done \cite{shuryak,gale}, the
effect of the rescattering of daughters on the experimentally measured
branching ratio has not previously been considered.

\subsection{Model Results}

\begin{figure}
\centerline{\psfig{figure=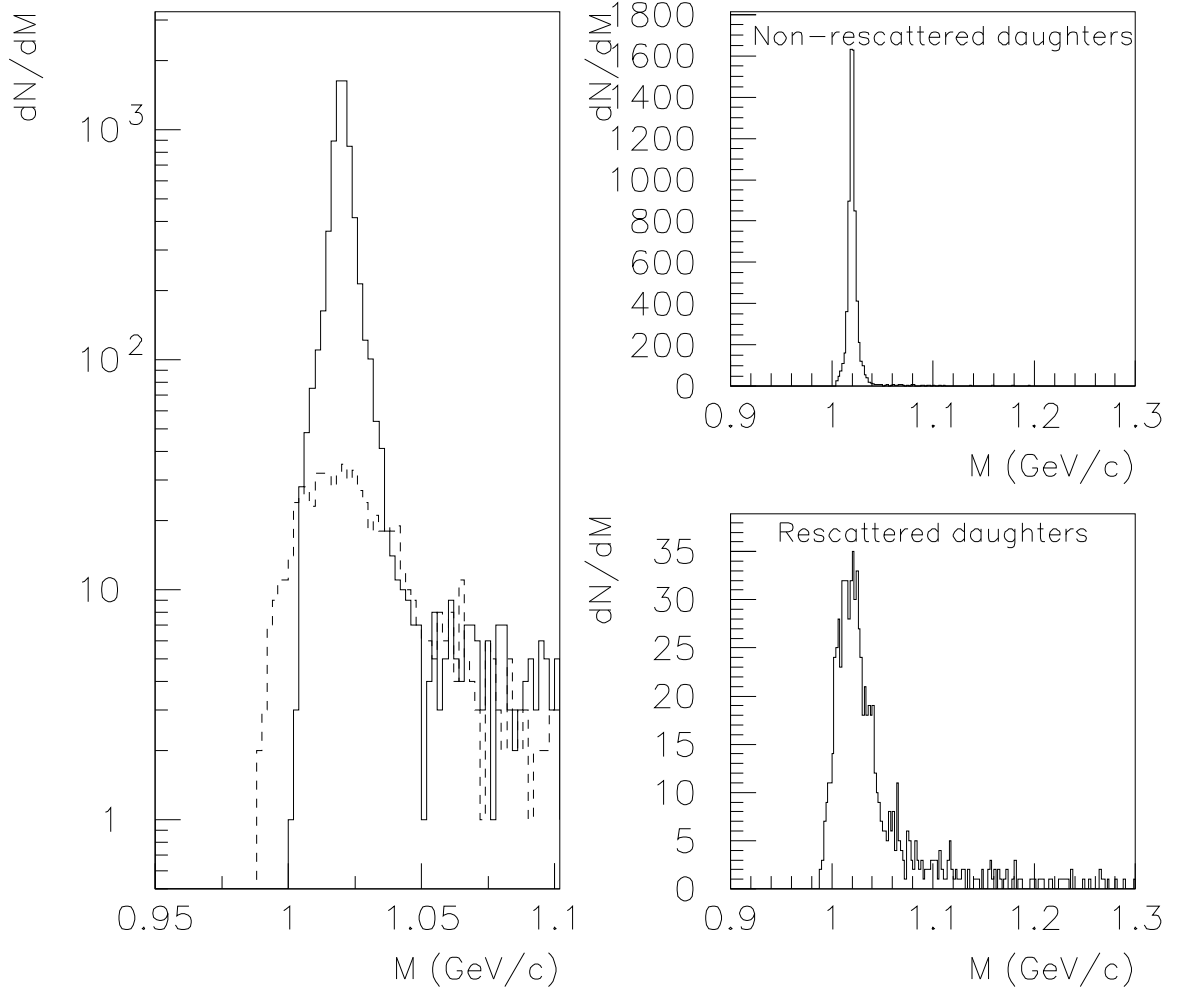,angle=0,height=17cm}}
\caption{Invariant mass distributions of reconstructed $\phi
\rightarrow K^+K^-$ in
RQMD.  The top right hand figure shows the reconstructed peak for
$\phi$'s with non-rescattered daughter kaons.  The bottom right figure
is for those $\phi$'s who had either one or both daughters rescatter
before leaving the collision zone.  The left hand figure shows an
overlay of these two plots.}
\label{fig:minv}
\end{figure}

To study the effect of hadronic scattering of daughters, we implemented
the hadronic cascade code
RQMD version 2.4 to describe the space time distribution of $\phi$'s and their
daughters.  In RQMD one can follow the history of all $\phi$'s that decay
throughout the collision along with their daughter kaons.  Upon
simulation of an event we determine the positions and momenta of
all kaons originating from $\phi$ decays.
Fig.~\ref{fig:minv} shows the invariant mass distribution of all kaon
pairs from  $\phi$ decays in simulated central Pb+Pb events at
158 GeV$\cdot$A/c beam energy.  The right hand figures show
$\phi$'s whose daughters escaped
from the collision zone without rescattering (top figure) and those
whose daughters did rescatter (bottom) yet still
escaped the collision zone as kaons.  The left hand
figure gives an overlay for comparison.  Since the $\phi$ lifetime is
comparable to, though larger than, the expected lifetime of a heavy ion
collision, most $\phi$'s do not decay in medium resulting in a tight
peak of the invariant mass ($M_{inv}$) with a width of $\sim$4 MeV.
However, $\sim 17$\% of $\phi$'s
decaying to two kaons have at least one daughter kaon that
rescatters and another (non-orthogonal) $\sim 17$\% of $\phi$'s have
at least one daughter that is absorbed.  The sum of
these two processes leads to $\sim 26$\% of decaying $\phi$'s with an
absorbed or rescattered daughter.
The $M_{inv}$ distribution of these kaons is much broader than the
original distribution; broad enough to escape detection, in particular
in experiments with low statistics or large combinatorial backgrounds.

Tables \ref{tab:kplus} and \ref{tab:kminus} itemize the channels
contributing to the first hadronic scattering of a daughter kaon from a
phi meson.  The dominant interactions for both $K^+$ and $K^-$
daughters proceed through
the $K^*$ resonance with a pion or through annihilation of the strange
quark by interaction with another kaon.  Nearly one-quarter of
all rescatterings occur through a variety of high mass resonances and
a sizeable fraction of the rescattered $K^-$ daughters proceed by a
reaction with a nucleon into a $\Sigma$ or other high mass strange
baryon.

\begin{figure}
\centerline{\psfig{figure=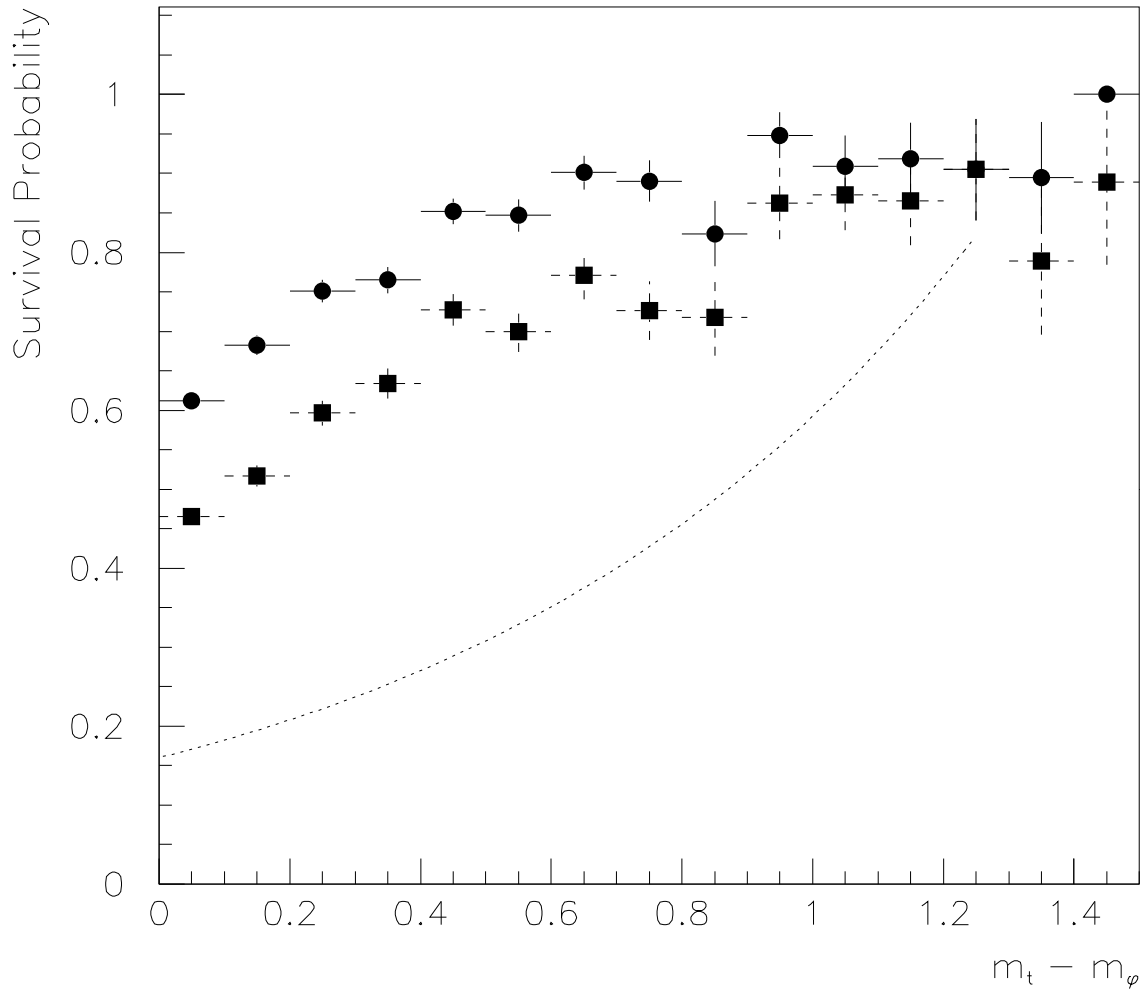,angle=0,height=17cm}}
\caption{The probability that a $\phi$ produced in
RQMD that decays to two kaons will have daughters that escape the
collision zone without rescattering as a function of $m_t$ (circles)
compared to experimental measurements of $\phi \rightarrow \mu \mu$
(NA50) to $\phi \rightarrow KK$ (NA49) corrected for the appropriate
branching ratio [5-7] (dotted line).  The squares represent
the maximal possible depletion of $\phi$ mesons as described in the
text.  All points corresond to those $\phi$'s with rapidity $|y|<1$.}
\label{fig:survive}
\end{figure}

The circles in
Fig.~\ref{fig:survive} display the $\phi$ survival probability, {\em
i.e.} the probability that neither of the daughter kaons rescatter, as a
function of $m_t$ for the rapidity region $|y|<1$.  The probability
decreases with
$m_t$, approaching $\sim 60$\% at the lowest $m_t$.  As a result,
the measured yield of $\phi$'s through the kaon
channel should be lower at low $m_t$ than that
measured in the dilepton channels.  The ratio of the yield of $\phi
\rightarrow K^+K^-$ to $\phi \rightarrow l^+l^-$ corrected by the
branching ratio should then approximately follow the circle symbols in
Fig.~\ref{fig:survive} if daughter rescattering is the sole mechanism
contributing to the difference.

\begin{figure}
\centerline{\psfig{figure=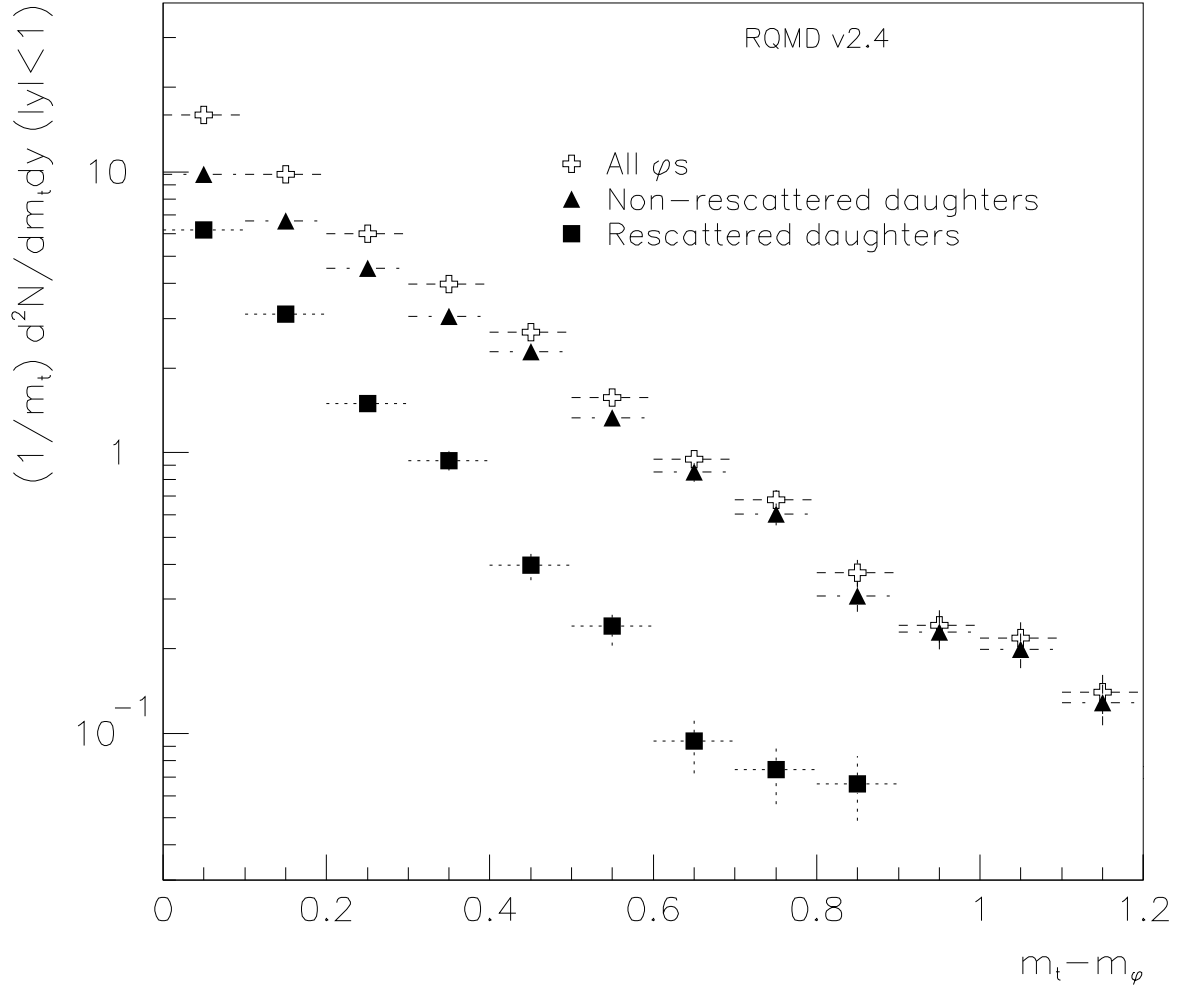,angle=0,height=17cm}}
\caption{The transverse multiplicity distributions of all $\phi$'s
that decay to two kaons in RQMD (crosses), as well as those those
whose daughters rescatter (squares) and those whose daughters do not
rescatter (triangles).  All multiplicity distributions have been
corrected for the $\phi$ branching ratio to $K^+K^-$.}
\label{fig:rescatt_phis}
\end{figure}

\begin{figure}
\centerline{\psfig{figure=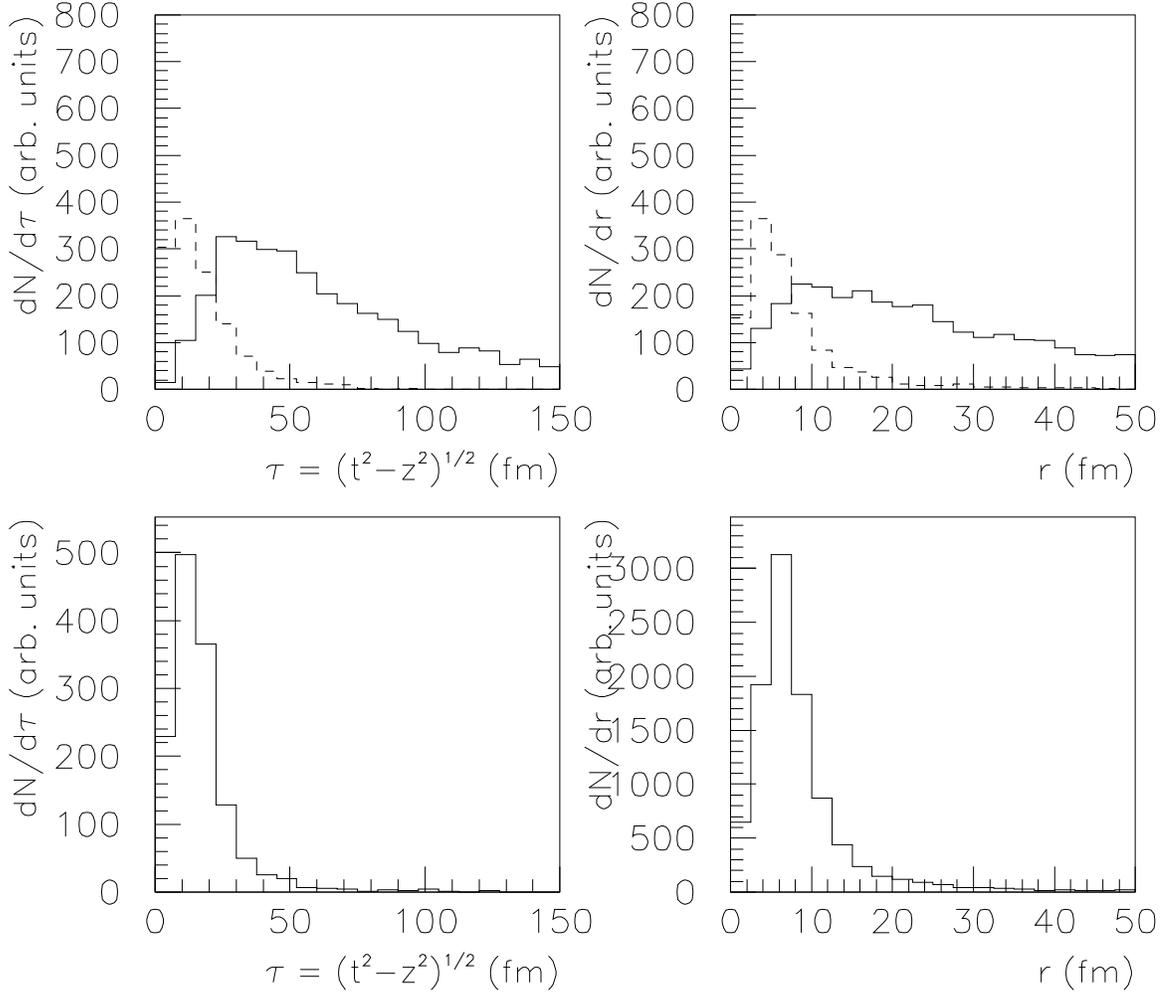,angle=0,width=17cm}}
\caption{The two top plots show the space time freeze out
distributions for $\phi$'s whose daughters rescattered (dashed) as
well as those whose daughters escaped the collision zone without
rescattering (solid).  The bottom two plots show the freeze out
distribution of all kaons from Pb+Pb collisions at the SPS.  $\tau$ is
the longitudinal proper time $\tau^2 = t^2 - z^2$ and r is the radial
position $r^2=x^2 + y^2$.}
\label{fig:space_time}
\end{figure}

Figs.~\ref{fig:rescatt_phis} and \ref{fig:space_time} show reconstructed
$\phi$ yields from RQMD as a function of $m_t$ and space-time,
respectively for $|y|<1$.  Fig.~\ref{fig:rescatt_phis} shows
$d^2N_\phi/m_t dm_t dy$ at mid rapidity for all $\phi$
mesons which decay in RQMD overlayed with the $m_t$ distribution
for those $\phi$'s whose daughters kaons do not rescatter and those 
$\phi$'s whose daughters do rescatter.  The
depletion of reconstructed $\phi$'s at low $m_t$ results in a
higher effective temperature of the $\phi$ meson at low $m_t$.  The
inverse slope for $0<m_t - m_\phi < 1$ GeV of
the original $\phi$'s is $T = 220 \pm 3$ MeV while for $\phi$'s
observed through the kaon decay channel RQMD predicts $T=242 \pm
4$ MeV.  The extracted yields and temperatures of $\phi$'s should
therefore be measureably different due to daughter rescattering.

The $m_t$ dependence of the rescattering effect upon
$\phi$'s is a reflection of the phase space freeze-out
distribution of particles in heavy ion collisions.
Fig.~\ref{fig:space_time} shows the freeze-out distribution in time and space
of the $\phi$'s from RQMD.  The left hand figure shows the freeze
out longitudinal proper time $\tau = \sqrt{t^2 - z^2}$ where $t$ and
$z$ are the freeze-out time and z-position, while the right hand figure
displays the radial freeze-out distribution.  The solid line
corresponds to those $\phi$'s whose daughters escaped unscathed from
the collision, while the dashed line shows the distribution for those
whose daughters did rescatter.  It is clear from the figure, and
intuitive, that those $\phi$'s with rescattered daughters
predominantly decayed at early times in the dense nuclear media,
during that part of the collision of particular interest for chiral
symmetry measurements.

\begin{figure}
\centerline{\psfig{figure=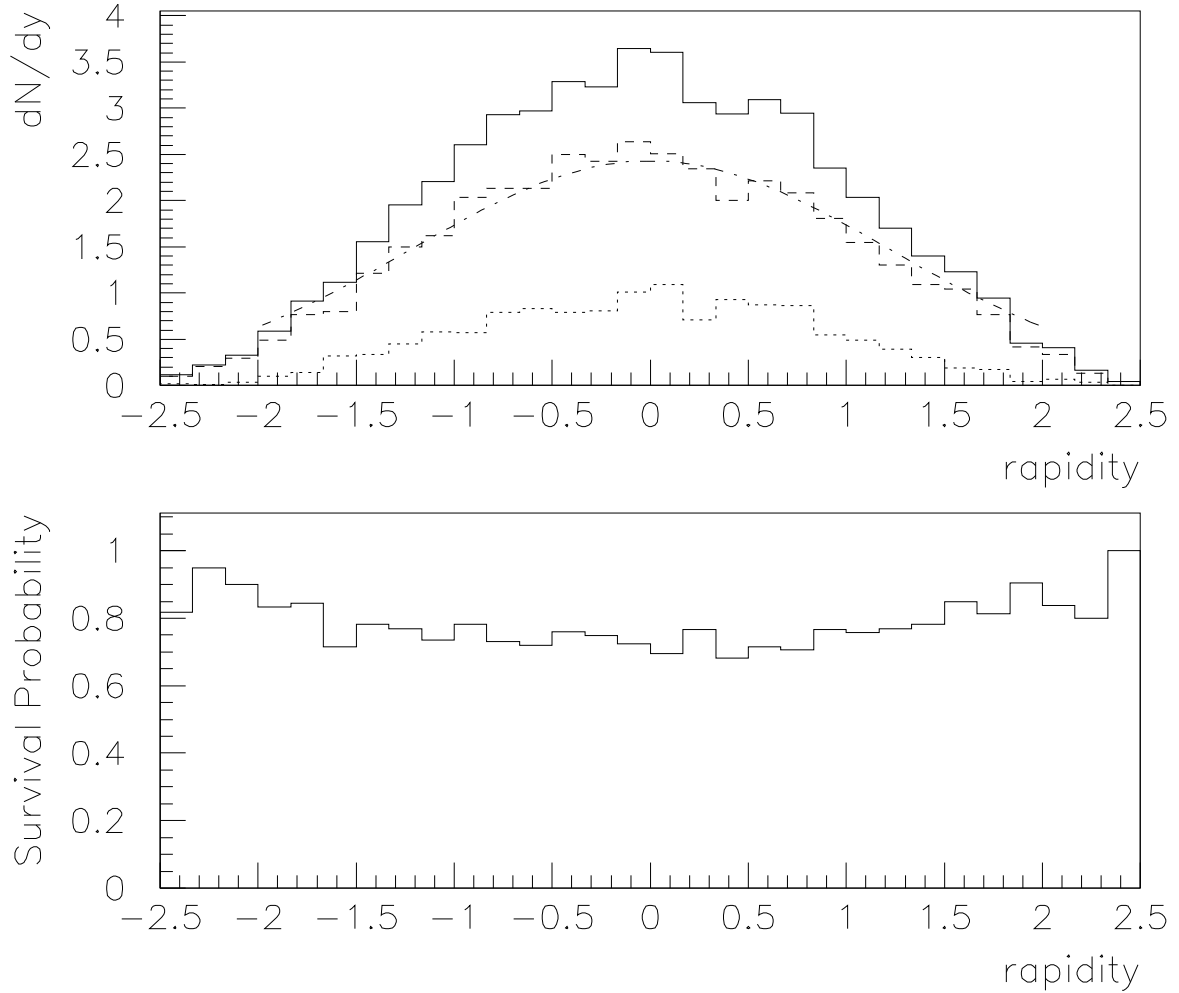,angle=0,width=17cm}}
\caption{Rapidity distributions of $\phi$'s decaying to $K^+$ $K^-$
corrected for the branching ratio from Pb+Pb collisions at
the SPS determined by RQMD v2.4.  The top plot shows the rapidity
distribution for all $\phi$'s (top), all $\phi$'s whose daughters were
not rescattered or absorbed (middle), and all $\phi$'s whose daughter
were rescattered or absorbed (bottom).  The curve corresponds to a fit
to the experimental data as described in the text. The bottom plot
displays the survival probability as a function of rapidity.}
\label{fig:rapidity}
\end{figure}

This picture is echoed in the rapidity distributions in
Fig.~\ref{fig:rapidity} where in the top plot we present the rapidity
of (1) all $\phi$'s, (2) those $\phi$'s whose daughter kaons escape
the collision zone unperturbed and (3) those $\phi$'s whose daughters
do not escape.  Here we see the approximately 25\% of all
$\phi$'s that are lost in the collision.  The effect of rescattering
is similar to that seen in the $m_t$ distribution as those $\phi$'s
closer to $y=0$ have a larger probability of being rescattered.  In
the bottom half of Fig.~\ref{fig:rapidity} we plot the probability of
survival of both daughter kaons which has a clear rapidity
dependence.  This dependence leads to a marginal widening of the
measured $\phi \rightarrow KK$ rapidity distribution relative to that
of the $\phi \rightarrow ll$ channel.

The conclusions from RQMD simulations are intuitive.  Those $\phi$'s
that decay at early times are more apt to have rescattered
daughter kaons due to the amount of hadronic matter through which they
must travel.  Further, rescattering and pressure build-up during
the collision implies that particles that freeze-out early in the
collision will have a softer transverse mass distribution than those
which freeze-out later \cite{ulrich}.  We then expect that the
rescattering effects at early times will be reflected in the
transverse mass distribution as confirmed in RQMD.

In order to compare these observations with experimental
measurements we note that RQMD only includes the imaginary part of the 
cross section in its calculations.  While Shuryak and Thorrsen
\cite{shuryak2} have shown that the real cross section for kaons in a
dense nuclear medium is much smaller than the imaginary cross section, 
we estimate here the maximal effect on the observed cross
section expected from the inclusion of the real part.

The addition of the real cross section can, at most, rescatter or 
absorb all kaons from $\phi$'s that decay within the freeze-out
volume.  Shown in the bottom two plots of Fig.~\ref{fig:space_time} is
the point of last rescattering for kaons from RQMD.  We define the
freeze-out volume for kaons to be that point within which 95\% of all
kaons have 
had their last interaction, which corresponds to $\tau \le 36.5$fm and
$r \le 20$fm.  The maximal effect then of adding the real cross
section to the RQMD calculation would be having all $\phi$'s that
decay within these ($\tau$,r) bounds be unreconstructable and lost
from the invariant multiplicity.

The square symbols in Fig.~\ref{fig:survive} show the result of
making such a drastic assumption.  In the lowest $m_t$ bin
approximately 25\% more $\phi$'s are depleted than in the imaginary
only calculation.  Note that the inclusion of the real part of the
cross section will only change the quantitative result from RQMD
slightly while the qualitative shape remains the same.

\subsection{Discussion}

Recently, two experiments at the SPS studying Pb+Pb collisions with a
beam energy of 158GeV/c have made preliminary reports of
$\phi$ measurements\cite{na50,na49,friese}.  Experiment NA50
\cite{na50} measured
$\phi \rightarrow \mu^+\mu^-$ at mid rapidity over the transverse mass
($m_t$) range $1.7 < m_t < 3.2$ GeV/c$^2$ while experiment 
NA49 \cite{na49,friese} reported a $\phi \rightarrow K^+K^-$ distribution
also at midrapidity but for $1 < m_t < 2.2$ GeV/c$^2$.
The reported $m_t$ inverse slopes are strikingly different; NA50 quotes
$T = 218 \pm 10$ MeV while NA49 finds $T = 305 \pm 15$ MeV.

Although, within the present accuracy, the data seem to be more
consistent in the $m_t$ range where they overlap, the extrapolated
yields are significantly different.  This points either towards a
drastic softening of the $m_t$ distribution with increasing $m_t$ or
distinctly different spectra reconstructed from $\phi \rightarrow KK$
and $\phi \rightarrow \mu\mu$.  The latter is in qualitative agreement
with the effect of rescattering of the decay kaons, depleting the low
$m_t$ region.  Quantitatively RQMD predicts a 17 MeV difference of the
slope, much smaller than the observed difference.  The curve in
Fig.~\ref{fig:survive} shows the ratio of the two extrapolated spectra,
$\phi \rightarrow KK / \phi \rightarrow \mu \mu$,
compared to the calculations from RQMD for the expected and maximal
effect of rescattered daughter kaons.  The curve is well below what
could be described by even the maximal daughter rescattering and we
conclude that this effect can not by itself describe the experimental data.

It is interesting to note, however, that RQMD does reproduce the
observed $\phi \rightarrow KK$ rapidity
distribution.  The line in Fig.~\ref{fig:rapidity} corresponds to the
experimentalists gaussian fit to their data in \cite{friese} with no
renormalization on our part.  This line corresponds quite nicely to
the RQMD curve for measured $\phi \rightarrow KK$ for those phi's
whose daughters did not rescatter.  The gaussian width and height of
the distribution from RQMD ($\sigma = .96 \pm .1$, $A = 2.45 \pm .46$)
are approximately consistent with the fit to the experimental data
($\sigma = 1.22 \pm .17$, $A = 2.43 \pm .15$) though the RQMD
distribution is not particularly well described by a gaussian.

Comparisons of $\phi \rightarrow KK$
with $\phi \rightarrow ll$ may be very informative if the spectra are
measured over the same range of $m_t$, with similar systematics.  If
the ratio of the $m_t$ spectra has the shape characterstic of
rescattering, the low $m_t$ dip in the $\phi \rightarrow KK/ll$ ratio
reflects the
amount of rescattering and therefore the time spent in the dense
hadronic phase.  This could help to clear up uncertainties about how
long the hadronic system interacts before freezing out
\cite{stock,wa98}.
Including effects of chiral symmetry restoration on kaon properties
may alter these arguments, but it is unlikely that both effects
will produce identical $m_t$ dependencies.

\subsection{Acknowledgements}

We would like to thank Dr.~H.~Sorge for the use of the RQMD code and
gratefully acknowledge helpful discussions with Dr.~E.~Shuryak.  We
further acknowledge the aid of Drs.~C.~Hoehne
and N.~Willis for aid in acquiring and interpreting the NA49 and NA50
data, respectively.

\begin{table}[t]
\centering
\begin{tabular}{||c|c||}
Interaction & Percentage \\ \hline \hline
$K^+ \pi \rightarrow K^*$ & 24\%  \\
$K^+ K^- \rightarrow \pi \pi$ or $ K K $ or $ f$ & 26\% \\
$K^+ K^0 \rightarrow \pi \eta $ or $ K^+ K^0$ & 20\% \\
$K^+ $ High Mass Resonance $ \rightarrow X$ & 27\% \\
All others & 3\% \\
\end{tabular}
\caption{The dominant channels in the rescattering of $K^+$ daughters
from $\phi$ decays in Pb+Pb collisions at SPS energies.}
\label{tab:kplus}
\end{table}

\begin{table}[tp]
\centering
\begin{tabular}{||c|c||}
Interaction & Percentage \\ \hline \hline
$K^- K^+ \rightarrow \pi \pi $ or $ \pi \rho $ or $ \pi \eta$ & 23\%  \\
$K^- K^0 \rightarrow \pi \eta $ or $ K K $ & 10\% \\
$K^- \pi \rightarrow K^* $ or $ K \pi$ & 21\% \\
$K^- n \rightarrow X$ & 8\% \\
$K^- p \rightarrow X$ & 17\% \\
$K^- $ High Mass Resonance $ \rightarrow X$ & 21\% \\
\end{tabular}
\caption{The dominant channels in the rescattering of $K^-$ daughters
from $\phi$ decays in Pb+Pb collisions at SPS energies.}
\label{tab:kminus}
\end{table}

\end{document}